\documentclass[aps, prd, twocolumn, nofootinbib, preprintnumbers, showpacs]{revtex4}
\usepackage{graphicx}
\usepackage{amsmath}
\usepackage{multirow}
\usepackage{bm}
\usepackage{color}
\def\fsl#1{\setbox0=\hbox{$#1$}                 
   \dimen0=\wd0                                 
   \setbox1=\hbox{/} \dimen1=\wd1               
   \ifdim\dimen0>\dimen1                        
      \rlap{\hbox to \dimen0{\hfil/\hfil}}      
      #1                                        
   \else                                        
      \rlap{\hbox to \dimen1{\hfil$#1$\hfil}}   
      /                                         
   \fi}                                         %

\begin{document}
\title{Dynamics behind the Quark Mass Hierarchy}
\date{\today}
\preprint{UWO-TH-09/2}
\author{Michio Hashimoto}
  \email{mhashimo@uwo.ca}
\author{V.A. Miransky}
  \email{vmiransk@uwo.ca}
\affiliation{
Department of Applied Mathematics, University of Western
Ontario, London, Ontario N6A 5B7, Canada}

\begin{abstract}
We introduce a new class of models describing the quark mass hierarchy. In 
this class, the dynamics primarily responsible for electroweak symmetry
breaking (EWSB) leads to the mass spectrum of quarks with no (or weak) isospin 
violation. Moreover, the values of these masses are of the order of the 
observed masses of the down-type quarks. Then, strong (although
subcritical) horizontal diagonal interactions for the $t$ quark plus
horizontal flavor-changing neutral interactions between different
families lead (with no fine tuning) to a realistic quark mass spectrum. 
In this scenario, 
many composite Higgs bosons occur. A concrete model
with the dynamical EWSB with the fourth family is described in detail.

\end{abstract}

\maketitle

\section{Introduction: Scenario}
\label{1}

The masses of quarks are~\cite{pdg}:
\begin{eqnarray}
&&  m_t=171.2 \pm 2.1 \mbox{ GeV}, \quad 
  m_b=4.20 \stackrel{\mbox{\scriptsize +0.17}}{\mbox{\scriptsize $-0.07$}}
  \mbox{ GeV}, \\
&&  m_c=1.27 \stackrel{\mbox{\scriptsize +0.07}}{\mbox{\scriptsize $-0.11$}}
  \mbox{ GeV}, \quad
  m_s=104 \stackrel{\mbox{\scriptsize +26}}{\mbox{\scriptsize $-34$}}
  \mbox{ MeV}, 
\end{eqnarray}
and
\begin{equation}
  m_u=1.5-3.3\mbox{ MeV}, \quad m_d=3.5-6.0\mbox{ MeV} \, . 
\end{equation}
The quark spectrum is characterized by the following striking features: (1)
There is a large hierarchy between quark masses from different families,
\begin{eqnarray}
 m_u/m_t \sim 10^{-5},\; m_u/m_c \sim 10^{-3}, \;
 m_c/m_t \sim 10^{-2}, \\
 m_d/m_b \sim 10^{-3},\; m_d/m_s \sim 10^{-2}, \;
 m_s/m_b \sim 10^{-1}\, .
\end{eqnarray}
(2) The isospin violation is also hierarchical: It is very strong in the 
third family, strong (although essentially weaker) in the second family, 
and mild in the first one:
\begin{equation}
 \frac{m_t}{m_b} \simeq  40.8, \quad \frac{m_c}{m_s}\simeq 11.5, \quad
 \frac{m_u}{m_d}=0.35-0.60 \, . 
\end{equation}
The origin of these features is still mysterious:
In the Standard Model (SM), 
it is required to introduce hierarchical yukawa couplings
by hand, e.g., $y_u/y_t = m_u/m_t \sim 10^{-5}$.
 
In this paper, we will introduce a new class of models describing the quark
mass hierarchy. One of our basic assumptions is the separation of
the dynamics triggering
the strong isospin violation in the third and second families from that
responsible for
the generation of the $W$ and $Z$ masses, i.e., electroweak 
symmetry breaking (EWSB).
The latter could be provided by one of the following known 
mechanisms: a) An elementary Higgs field (or fields).
b) A modern version of the technicolor (TC) scenario  
(for recent reviews, see Ref.~\cite{TC}). 
c) At last, it could be a dynamical Higgs mechanism with a Higgs doublet (or
doublets) composed of $t^{\prime}$ and $b^{\prime}$ quarks of the fourth
family~\cite{4family,Burdman:2007sx}.

\begin{figure}[t]
  \centering
  \resizebox{0.25\textheight}{!}{\includegraphics{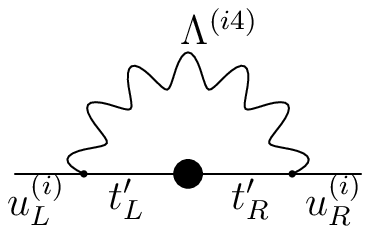}}\\[8mm]
  \resizebox{0.25\textheight}{!}{\includegraphics{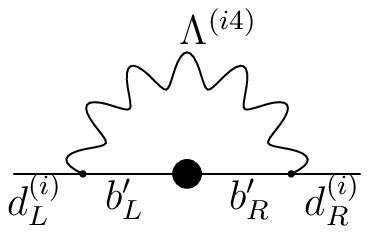}}
  \caption{FCN interactions of the up- and down-quark sectors.
           Here $u^{(1,2,3)}=u,c,t$ and $d^{(1,2,3)}=d,s,b$, 
           respectively. $\Lambda^{(i4)}$ are masses of exchange
           vector particles. \label{tp-ui}}
\end{figure}

We assume that 
the dynamics primarily responsible for the EWSB 
leads to the mass spectrum of quarks
with no (or weak) isospin violation. 
{\it Moreover, we assume that 
the values of these masses are of the order of the observed masses of the 
down-type quarks.} In the case of an elementary Higgs field
(or fields), they are provided by the conventional yukawa interactions.
In the case of the dynamical Higgs mechanism, in order to generate these 
masses, one should use flavor-changing-neutral (FCN) interactions: 
the extended technicolor (ETC)~\cite{ETC} in the case of the TC scenario,
and the horizontal interactions between the 4th family and the first
three ones in the case of the scenario with the fourth 
family (see Fig.~\ref{tp-ui}).

Of course, such interactions are restricted by
the $K^0$-$\bar{K}^0$ mixing, for example, and thus for light quarks
it is required to introduce heavy exchange vector particles, 
say, with the masses of order 1000 TeV.
Such heavy particles can be a natural source for producing 
small yukawa coupling constants for light quarks.
For heavier quarks, we introduce lighter vector particles.
 
\begin{figure}[t]
  \centering
  \resizebox{0.25\textheight}{!}{\includegraphics{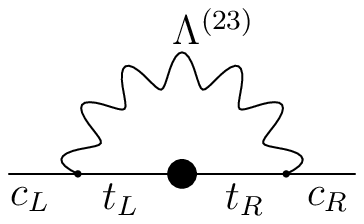}}\\[10mm]
  \resizebox{0.25\textheight}{!}{\includegraphics{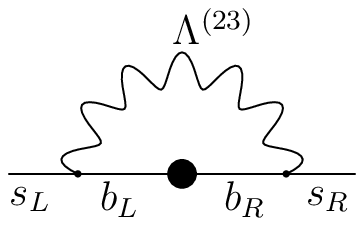}}
  \caption{FCN interactions for the second family. \label{c-t}}
\end{figure}

The second (central) stage is introducing the horizontal 
interactions for the quarks in the first three families (this stage
is essentially the same for the three EWSB mechanisms mentioned above.)
First, following the idea in the model
of Mendel et al.~\cite{Mendel:1991cx,book}, 
we utilize strong 
(although {\it subcritical}) diagonal horizontal interactions 
for the top 
quark which lead to the observed ratio $\frac{m_t}{m_b} \simeq 40.8$.
The second step is introducing the equal strengths 
horizontal FCN interactions between the $t$ and $c$ quarks
and the $b$ and $s$ ones in order to get the observed ratio 
$m_c/m_s \simeq 11.5$
in the second family (see Fig.~\ref{c-t}).
As will be shown in Sec.~\ref{2.2}, these
interactions can naturally provide such a ratio indeed. 

Because of a smallness of the mixing angles for
quarks from the different families, neglecting the family mixing 
in the dynamics responsible for generating the quark masses in the
second and third families is a reasonable approximation.
Concerning the mild isospin
violation in the first family, it should be studied together with
the effects of the family mixing, reflected in the 
Cabibbo-Kobayashi-Maskawa (CKM) matrix. The latter will be considered in
Sec.~\ref{2.3}.

Thus, in the present scenario, beside the EWSB interactions, 
the dominant 
dynamics responsible for the form of the mass spectrum of quarks is
connected with the diagonal horizontal interactions for the third family and
the horizontal FCN interactions between the second and third ones.  
The signature of this scenario is the appearance of
composite Higgs bosons (resonances) composed of the quarks and
antiquarks of the 3rd family (see Sec.~\ref{2.4})

The main source of
the isospin violation in this approach
is the strong top quark interactions. On the other
hand, because these interactions are subcritical, the top quark plays a minor
role in electroweak symmetry breaking. This point distinguishes this
scenario from the top quark condensate 
model~\cite{Miransky:1988xi,Nambu,Marciano:1989xd,Bardeen:1989ds}.    

Two comments are in order.
(i) As will be shown below, the characteristic feature in this class of the
models is the absence of fine tuning:
What could be called fine tuning for the nearcritical coupling of the
$t$ quark (1 part in $10^2$) is just a reflection of a ``unnaturally'' large
isospin violation in the third family, $m_b/m_t \simeq 2.5 \times 10^{-2}$.
(ii) In this paper, we will concentrate on studying the mass 
spectrum 
of quarks. For a discussion concerning the extension of the present approach
for the description of lepton masses, see Sec.~\ref{4} below.

\section{Model}
\label{2}

In this section, the dynamics for generating the 
quark mass hierarchy
will be described in detail. Henceforth we will concentrate on a
model of the dynamical EWSB with the fourth family~\cite{4family}.
However, we will also comment on the modifications (if any) for
both the scenario with elementary Higgs fields responsible for
the EWSB and the TC scenario.

\subsection{Electroweak symmetry breaking dynamics and isospin 
symmetric quark masses}
\label{2.1}

The first stage is generating the masses with no (or weak) isospin
violation and of the order of the observed masses of the down-type
quarks. As was pointed in the Introduction, in the present approach,
the EWSB dynamics is responsible for that. 
It is straightforward to produce such masses  both in the case of the
scenario with
elementary Higgs fields (through yukawa interactions) and in the TC
one (through ETC interactions). 

Let us now describe this stage in the scenario of the dynamical EWSB
with the fourth family~\cite{4family}.
The masses of the 4th family  quarks are constrained as~\cite{pdg}
\begin{equation}
  m_{b'} > 199\mbox{ GeV}, \quad   m_{t'} > 256\mbox{ GeV} \, .
\end{equation}
Note that if the mixing angles between the 4th family and the rest ones
are extremely small, $b'$ and $t'$ quarks
behave like long-lived charged massive particles. In 
this case the constraints are $m_{b'} > 190$ GeV and
$m_{t'} > 220$ GeV~\cite{Acosta:2002ju}.

At the composite scale $\Lambda^{(4)}$, the 4th family quarks 
$t'$ and $b'$ condense and thereby they break the electroweak symmetry.
By using the Pagels-Stokar (PS) formula~\cite{PS,book}, we can estimate
the corresponding decay constants,
\begin{subequations}
\label{PS}
\begin{eqnarray}
  v_{t'}^2 &=& \frac{N}{8\pi^2}m_{t'}^2
  \ln\left(1+\frac{(\Lambda^{(4)})^2}{m_{t'}^2}\right), 
  \label{v-tp} \\
  v_{b'}^2 &=& \frac{N}{8\pi^2}m_{b'}^2
  \ln\left(1+\frac{(\Lambda^{(4)})^2}{m_{b'}^2}\right), 
  \label{v-bp}
\end{eqnarray}
\end{subequations}
with
\begin{equation}
  v_{t'}^2 + v_{b'}^2 = v^2 , 
  \label{v2-tp-vp}
\end{equation}
where $N=3$ and $v=246$ GeV.
The constraint of the $T$-parameter suggests 
that $m_{t'} \simeq m_{b'}$ is favorable and thereby
$v_{t'} \simeq v_{b'}$ follows.
Note that the masses of $t'$ and $b'$ are essentially determined 
through the PS formula (\ref{PS}) 
when the value of $\Lambda^{(4)}$ is fixed.

In order to obtain almost correct masses for the down-type quarks, 
\begin{equation}
  m_0^{(3)} \sim \mbox{1 GeV}, \;   m_0^{(2)} \sim \mbox{100 MeV}, \;
  m_0^{(1)} \sim \mbox{1 MeV} , 
\end{equation}
we introduce the following horizontal 
FCN interactions (see Fig.~\ref{tp-ui}):
\begin{equation}
  t'-u^{(i)}-\Lambda^{(i4)}, \quad b'-d^{(i)}-\Lambda^{(i4)} ,  
\end{equation}
where $i=1,2,3$ and $u^{(1,2,3)}=u,c,t$ and $d^{(1,2,3)}=d,s,b$, 
respectively.
These one-loop contributions yield
\begin{equation}
  m_0^{(i)} \simeq
  \frac{C_2 g_{t'u^{(i)}}^2}{4\pi^2}
  \frac{(\Lambda^{(4)})^2}{(\Lambda^{(i4)})^2} m_{t'}
  \simeq
  \frac{C_2 g_{b'd^{(i)}}^2}{4\pi^2}
  \frac{(\Lambda^{(4)})^2}{(\Lambda^{(i4)})^2} m_{b'},
\label{m}
\end{equation}
where $C_2$ represents the quadratic Casimir invariant and 
we took into account that the dynamical running 
$m_{t'}$ and $m_{b'}$ masses rapidly decrease above the scale
$\Lambda^{(4)}$ (if these masses are not sharply cutoff at 
$\Lambda^{(4)}$, there can appear log$(\Lambda^{(4)})$ 
factors in Eq.~(\ref{m}), as in 
QCD~\cite{Lane:1974he,Miransky:1983vj}). 

In order to obtain the hierarchical masses $m_0^{(1,2,3)}$,
we assume 
\begin{equation}
 (\Lambda^{(14)})^2 \gg (\Lambda^{(24)})^2 \gg (\Lambda^{(34)})^2 \gg 
 (\Lambda^{(4)})^2. 
\end{equation}
We may expect 
$C_2 g_{t'u^{(i)}}^2 \simeq C_2 g_{b'd^{(i)}}^2 \sim {\cal O}(1)$.
Then, at this stage, the mass spectrum of quarks is isospin symmetric.
The running masses are essentially equal to the constants $m_0^{(i)}$
up to the scale of $\Lambda^{(i4)}$ $(i=1,2,3)$.
Above $\Lambda^{(i4)}$, they rapidly, 
as $1/q^2$, decrease ($q$ is the momentum
of the running masses).

In order to get the appropriate numbers, 
the scales should be determined by
\begin{equation}
  \eta_{t'(b')}^{(i)} \equiv 
  \frac{C_2 g_{t'u^{(i)}(b'd^{(i)})}^2}{4\pi^2}
  \frac{(\Lambda^{(4)})^2}{(\Lambda^{(i4)})^2}
  \simeq \frac{m_0^{(i)}}{m_{t'(b')}}.
\end{equation}

\subsection{Horizontal interactions as the source of the isospin violation 
in the quark masses}
\label{2.2}

The second (central) stage in the present scenario is introducing the 
horizontal interactions for the three known fermion families. Note that this 
stage is
essentially identical for the scenarios with the different EWSB dynamics: 
elementary Higgs fields, TC, and the fourth family.  

Let us start from the description of the dynamics generating 
the large top quark mass. At energy scales less
than the mass of a horizontal vector boson 
$\Lambda^{(3)} \sim \Lambda^{(34)}$,
the corresponding horizontal interactions can be presented by the
four-fermion Nambu-Jona-Lasinio (NJL) ones.
We apply strong (although subcritical) dynamics 
for the horizontal diagonal interactions
for the $t$ quark. The isospin symmetric
mass $m^{(3)}_0$, 
introduced in Sec. \ref{2.1}, plays the role of a
bare mass with respect to these interactions. The solution of the
Schwinger-Dyson equation for the $t$ quark propagator leads to the
following mass $m_t$ \cite{Mendel:1991cx,book}:
\begin{equation}
  m_{t} \simeq \frac{1}{\Delta g_t} m_0^{(3)} , 
\end{equation}
where $\Delta g_q$ denotes the difference of 
the critical coupling and 
the (normalized) dimensionless NJL one for a $q$ quark, so that 
\begin{equation}
 \Delta g_t \simeq \frac{m_0^{(3)}}{m_t} \sim 6 \times 10^{-3},
\end{equation}
where we used $m_t=171.2$ GeV and $m_0^{(3)}=1$ GeV.
For the bottom quark, it should be $\Delta g_b \sim {\cal O}(1)$.
In any case, it is required,
\begin{equation}
  \Delta g_b - \Delta g_t \simeq \frac{m_0^{(3)}}{m_b},
  \label{diff-t-b}
\end{equation}
where we ignored $m_0^{(3)}/m_t$ because of $m_t \gg m_b$. 
Concrete models for obtaining 
such a isospin symmetry breakdown in the third family are described
in Appendix~\ref{A}.

Let us now turn to the generation of the realistic masses for the second 
family. We assume that there exist FCN interactions 
between the $t$ and $c$ quarks and similarly between the $b$ and $s$ ones
(see Fig.~\ref{c-t}),
\begin{equation}
  t-c-\Lambda^{(23)}, \quad b-s-\Lambda^{(23)} \, . 
\end{equation}
These one-loop diagrams yield the following masses for
charm and strange quarks:
\begin{equation}
  m_c = m_0^{(2)} + \eta^{(23)}_t m_t, \quad
  m_s = m_0^{(2)} + \eta^{(23)}_b m_b, 
\end{equation}
where $m^{(2)}_{0} \sim 100$ MeV is the isospin symmetric mass
for the second family (see Sec. \ref{2.1}), and $\eta^{(23)}_{t,b}$ are
\begin{equation}
  \eta^{(23)}_{t (b)} \equiv 
  \frac{C_2 g_{tc (bs)}^2}{4\pi^2}
   \frac{(\Lambda^{(34)})^2}{(\Lambda^{(23)})^2}
\end{equation}
for $\Lambda^{(23)} \gg \Lambda^{(34)}$.

As described above, 
the ratio $m_b/m_t \simeq 1/40$ 
is obtained via the near-critical dynamics in this model.
Now, taking $m_0^{(2)} = 100$ MeV and
$\eta^{(23)}_{t} = \eta^{(23)}_{b} =1/100$, we get
\begin{eqnarray}
  m_c &=& 100\mbox{ MeV} + m_t/100 \sim \mbox{1 GeV}, \\
  m_s &=& 100\mbox{ MeV} + m_b/100 \sim \mbox{140 MeV} \, .
\end{eqnarray}
In this way, we can obtain the correct mass enhancement
for the charm quark via the large $m_t$.
{\it Let us emphasize that
the presence of the isospin symmetric mass
$m_0^{(2)} \sim \mbox{100 MeV} \sim m_s$ is
crucial here: 
with $m_0^{(2)} \ll 100$ MeV,
the ratio $m_s/m_c$ would be close to $m_b/m_t$.}

As to the horizontal FCN gauge bosons which couple to 
the quarks of the 1st and 2nd families, we assume that
they are very heavy,
\begin{equation}
  c-u-\Lambda^{(12)}, \quad s-d-\Lambda^{(12)} ,  
\end{equation}
with $\Lambda^{(12)} \gtrsim {\cal O}(1000\mbox{ TeV})$.
As a result, their contributions 
to the masses of the $u$ and $d$ quarks are very small.

\subsection{The CKM mass matrix}
\label{2.3}

\begin{figure}[t]
  \centering
  \resizebox{0.25\textheight}{!}{\includegraphics{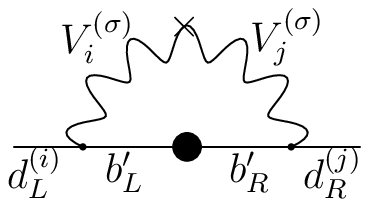}}
  \caption{FCN interactions with a gauge boson mixing.} 
The parameter $\sigma =1, 2$ is described in text. 
\label{mass-mixing}
\end{figure}

So far we have neglected the family mixing effects. Because the
mixing angles between quarks from the different families are small,
such an approach can be considered as a reasonable approximation 
for the description of
generating quark masses in the second and third quark families.
Here we will turn to the structure of the CKM mass matrix.

Recall that the number of the CP phases is three in the
4th family quark model, whereas the three generation model has
only one CP phase~\cite{Kobayashi:1973fv}.
This can offer richer phenomenology, for example, in the $B$ physics.
In this paper, however, we ignore
the CP violation and concentrate on the family mixing effects.

There are several approaches to this problem:
(1) Mass texture ans\"{a}tze
(for example, the Fritzsch-type mass matrix~\cite{Fritzsch:1977vd},
the democratic family mixing, etc.).
(2) The Froggatt-Nielsen mechanism~\cite{Froggatt:1978nt}.
(3) Dynamical approaches, e.g., ETC models~\cite{Appelquist:2003hn},
the top loops mechanism~\cite{Dobrescu:2008sz}, etc.. 
We will employ a modification 
of the dynamical approach in Ref.~\cite{Appelquist:2003hn} that is 
appropriate for the model with the 4th family.

Let us start from the down-type quark masses.
We assume that

\begin{enumerate}
\item There exist horizontal FCN interactions
with a mixing of $V_{i}^{(1)}$ and $V_{j}^{(1)}$ gauge bosons related
to two different families $i$ and $j$, one of which is the first one
(see Fig.~\ref{mass-mixing}).
We further assume that the values of all the relevant parameters 
(the masses of $V_{i}^{(1)}$ and $V_{j}^{(1)}$, 
and the gauge boson mixing parameters)
are around the scale $\Lambda^{(14)}$.
In this case, we obtain naturally a universal mass $m_{\rm off}^{(1)}$
with $m_{\rm off}^{(1)} \sim m_d$.
\item Similarly, when neither $i$ nor $j$ are $1$,
there exist horizontal FCN interactions with a mixing for another
set of $V_{i}^{(2)}$ and $V_{j}^{(2)}$ gauge bosons. In this case,
the values of all the relevant parameters are assumed to be around
$\Lambda^{(24)}$. This leads to a universal mass
$m_{\rm off}^{(2)} \sim m_s$.
\end{enumerate}

We can then explicitly write the mass matrix $M_D$ for
the down-type quark as
\begin{equation}
  M_D =
  \left(\begin{array}{cccc}
    m_d & \xi_1 m_d & \xi_1 m_d & \xi_1 m_d \\
    \xi_1 m_d & m_s & \xi_2 m_s & \xi_2 m_s \\
    \xi_1 m_d & \xi_2 m_s & m_b & \xi_2 m_s \\
    \xi_1 m_d & \xi_2 m_s & \xi_2 m_s & m_{b'}
  \end{array}\right) \, .
\end{equation}
The parameters $\xi_{1,2}$ will be determined by
$|V_{us}|$ and $|V_{cb}|$ later.
As to the diagonal mass terms, the values of 
$m_s$, $m_b$, and $m_{b'}$ are almost the same as the mass eigenvalues, 
whereas it is required to adjust numerically the value of $m_d$
in order to obtain the correct mass eigenvalue for the down quark.

For the up-type quarks, the mass matrix has a similar
structure with the replacement of $m_d, m_s, m_b, m_{b'}$ by
$m_u, m_c, m_t, m_{t'}$, respectively.

Since the mass matrix $M_D$ is symmetric,
it can be diagonalized by a single orthogonal matrix $D_L$.
Similarly, the up-type quark mass matrix can be diagonalized
by an orthogonal matrix $U_L$.
The $4 \times 4$ CKM matrix $V_{CKM}^{4 \times 4}$ is
given by
\begin{equation}
  V_{CKM}^{4 \times 4} = U_L^\dagger D_L \, .
\end{equation}

Noting that $m_d \ll m_s \ll m_b \ll m_{b'}$,
we approximately obtain the matrix $D_L$ as
\begin{widetext}
\begin{equation}
  D_L \simeq
  \left(\begin{array}{cccc}
    1-\frac{\xi_1^2}{2}\left(\frac{m_d}{m_s}\right)^2 &
    \xi_1 \frac{m_d}{m_s} & \xi_1 \frac{m_d}{m_b} &
    \xi_1 \frac{m_d}{m_{b'}} \\
   -\xi_1 \frac{m_d}{m_s} &
    1-\frac{\xi_1^2}{2}\left(\frac{m_d}{m_s}\right)^2 &
    \xi_2 \frac{m_s}{m_b} & \xi_2 \frac{m_s}{m_{b'}} \\
   -\xi_1 \frac{m_d}{m_b} &-\xi_2 \frac{m_s}{m_b} &
    1 & \xi_2 \frac{m_s}{m_{b'}} \\
   -\xi_1 \frac{m_d}{m_{b'}} & -\xi_2 \frac{m_s}{m_{b'}} &
   -\xi_2 \frac{m_s}{m_{b'}} & 1
  \end{array}\right),
\end{equation}
\end{widetext}
where we took into account that the quadratic term $m_d^2/m_s^2 \sim O(0.01)$.

On the other hand, since numerically
$m_u/m_c \ll m_d/m_s$, $m_u/m_t \ll m_d/m_b$,
and $m_c/m_t \ll m_s/m_b$,
we can neglect the off-diagonal entries of $U_L$
in the $3 \times 3$ part of the CKM matrix. 
Then we get:
\begin{eqnarray}
 && |V_{ud}| \simeq |V_{cs}| \simeq
    1-\frac{\xi_1^2}{2}\left(\frac{m_d}{m_s}\right)^2,\\
 && |V_{tb}| \simeq 1, \\
 && |V_{us}| \simeq |V_{cd}| \simeq \xi_1 \frac{m_d}{m_s},\\
 && |V_{ub}| \simeq |V_{td}| \simeq \xi_1 \frac{m_d}{m_b},\\
 && |V_{cb}| \simeq |V_{ts}| \simeq \xi_2 \frac{m_s}{m_b} \, .
\end{eqnarray}
The relation $|V_{ub}/V_{us}|=m_s/m_b=0.02$ is noticeable.
Note that the PDG value is
$|V_{ub}/V_{us}|=3.93 \times 10^{-3}/0.2255=0.0174$~\cite{pdg}.

By using $|V_{us}| = 0.23$ and $|V_{cb}|=0.04$~\cite{pdg},
we fix the values of $\xi_{1,2}$,
\begin{equation}
  \xi_1 = \frac{23}{m_d\mbox{(MeV)}}, \quad
  \xi_2 = 0.04 \times \frac{m_b}{m_s}=2 \, .
  \label{xi}
\end{equation}
With these values of $\xi_1$ and $\xi_2$,
and the masses of quarks
for $D_L$ and $U_L$, we thereby
obtain the $4 \times 4$ CKM matrix:
\begin{equation}
  V_{CKM}^{4 \times 4} =
  \left(\begin{array}{cccc}
  0.97 & 0.23 & -0.006 & 0.00009 \\
  -0.23 & 0.97 & -0.04 &  -0.008 \\
  -0.003 & 0.04 & 1.0 & 0.02 \\
  -0.002 & 0.007 & -0.02 & 1.0
  \end{array}\right),
  \label{example}
\end{equation}
where we used $m_{t'}=m_{b'}=300$ GeV, which is responsible
only for the 4th column and row.
Actually, the values of $|V_{ud}|$, $|V_{cs}|$, $|V_{tb}|$,
$|V_{cd}|$, and $|V_{ts}|$ are ``correct''~\cite{pdg}.
Although our $|V_{ub}|=0.006$ and $|V_{td}|=0.003$ are
a bit different from the PDG values~\cite{pdg},
the orders, $|V_{ub}| \sim |V_{td}| \sim O(0.001)$, are correct.

The 4th generation mixing terms are approximately given by
\begin{equation}
  |V_{t' d}| \simeq \xi_1 \frac{m_d}{m_s} \cdot \xi_2 \frac{m_c}{m_{t'}}
  \sim 0.23 \times \xi_2 \frac{m_c}{m_{t'}} \sim {\cal O}(10^{-3}) ,
  \label{vtpd}
\end{equation}
and
\begin{equation}
  |V_{t' s}| \simeq |V_{t' b}| \simeq \xi_2 \frac{m_c}{m_{t'}}
  \sim {\cal O}(10^{-2})\,.
\end{equation}
Thus the contributions of $t'$ to the $B^0$--$\bar{B}^0$ mixing
are roughly proportional to
$m_{t'}^2 |V_{t'd}^* V_{t'b}|^2 \sim m_c^4/m_{t'}^2 \times 10^{-2}$
for $B_d$ and $m_{t'}^2 |V_{t's}^* V_{t'b}|^2 \sim m_c^4/m_{t'}^2$
for $B_s$.
On the other hand, the corresponding SM contributions are
proportional to $m_{t}^2 |V_{td}^* V_{tb}|^2 \simeq 6.4 10^{-5} m_t^2$
and $m_{t}^2 |V_{ts}^* V_{tb}|^2 \simeq 1.6 10^{-3} m_t^2$,
respectively.
Therefore the 4th generation contributions are negligible.
Similarly, the processes $b \to s \gamma$ and
$Z \to b\bar{b}$ are also suppressed.

Although the dynamics underlying the CKM matrix is still far from
being completely understood, it is noticeable that
by using a simple extension of
the mechanism for producing the quark masses used
in Secs.~\ref{2.1} and \ref{2.2}, the essential features of
the CKM matrix can be extracted.

\subsection{Composite Higgs bosons}
\label{2.4}

In this scenario, there potentially appear many composite
Higgs bosons (compare with Refs.~\cite{Mendel:1991cx,book,Hill:1994hp}).
In the scenario with the 4th family quarks,
the masses of the bound states of the $t'$ and $b'$ quarks should be
of the order of the EWSB scale.
Since we consider the condensation both of the $t'$ and $b'$,
there appear at least two composite Higgs doublets.
For the 3rd family, we may estimate the mass of
the top-Higgs doublet (resonance) $\phi_t$ via 
the NJL relation~\cite{Appelquist:1991kn,Mendel:1991cx,book}:
\begin{equation}
  M_{\phi_t} \sim
  \Lambda^{(3)} 
  \left(\frac{2\Delta g_t}{\ln \frac{1}{2\Delta g_t}}\right)^{1/2}
  \sim 0.05 \Lambda^{(3)}, 
\label{thiggs}
\end{equation}
where we used $\Delta g_t \sim 6 \times 10^{-3}$.
For the bottom-Higgs resonance $\phi_b$, 
it should be $M_{\phi_b} \sim \Lambda^{(3)}$, i.e., it is 
very heavy and unstable. 
Note that the quark structures of the composites $\phi_t$ and $\phi_b$ are
$\phi_t \sim (\Lambda^{(3)})^{-2}\, \bar{t}_R (t, b)_L$ and
$\phi_b \sim (\Lambda^{(3)})^{-2}\, \bar{b}_R (b,-t)_L$, respectively.

Note that
in the case of the scenario with elementary Higgs fields responsible for
the EWSB, there should appear (beside the elementary Higgs fields)
at least one composite Higgs resonance $\phi_t$. 
In the TC scenario, such a Higgs resonance exists in addition to 
technihadrons.

Since we assume that the scales $\Lambda^{(1)}$ and $\Lambda^{(2)}$ (related
to the 1st and 2nd families) are very large, the corresponding Higgs
composites should be very heavy and unstable, and therefore they are
irrelevant for the electroweak dynamics.

\section{Phenomenological analysis}
\label{3}

In this section, we describe a phenomenology in the simplest
model with the 4th family of the class described in Sec.~\ref{2}.
In this model, the scale $\Lambda^{(3)}$ is assumed to be sufficiently
large, such that the mass $M_{\phi_t}$ (\ref{thiggs})
is much heavier than
the masses of the Higgs doublets composed of the $t'$ and $b'$.
Otherwise, the mass of the top-Higgs field $\phi_t$ would be also
of the order of the EWSB scale.
In that case, there appear three relevant Higgs doublets.
This interesting possibility will be considered elsewhere.

For $v_{t'}=v_{b'}$, the PS formula (\ref{PS}) yields
$m_{t'} = m_{b'} \sim 0.3$ TeV with $\Lambda^{(4)}=10$ TeV. 
More precisely, by using the RGE's~\cite{Hill:1985tg}
with the compositeness conditions~\cite{Bardeen:1989ds,Luty:1990bg},
we obtain
\begin{equation}
  m_{t'} = 0.292\mbox{ TeV}, \quad  m_{b'} = 0.291\mbox{ TeV} , 
\end{equation}
which gives the $T$-parameter contribution $T_f = 10^{-5}$.
Smaller $\Lambda^{(4)}$ provides larger $m_{t'}$ and $m_{b'}$
with relaxing the cost of the fine-tuning, 
due to a combination of the gap equation and the PS formula,
\begin{equation}
  \frac{v^2}{(\Lambda^{(4)})^2} =
  \frac{N}{8\pi^2}\left(\,
  2-\frac{1}{g_{t'}^{\rm eff}}-\frac{1}{g_{b'}^{\rm eff}}\,\right)
  \simeq \frac{N}{4\pi^2}\left(\,1-\frac{1}{g_{t'}^{\rm eff}}\,\right),
\end{equation}
where we used near-equality for 
the effective dimensionless NJL couplings:
$g_{b'}^{\rm eff} \simeq g_{t'}^{\rm eff}$,
because $m_{t'} \simeq m_{b'}$.

As to the masses of the Higgs bosons composed of $t'$ and $b'$,
we take the mass $M_A$ of the CP odd Higgs as a free parameter.
The rest masses are determined through 
the RGE's~\cite{Hill:1985tg,Luty:1990bg}.
For example, we may take
\begin{equation}
  M_A = 0.30, 0.40, 0.50, 0.60 \mbox{ TeV},
\end{equation}
and in this case, we obtain the charged and CP even Higgs masses,
\begin{eqnarray}
  M_{H^\pm} &=& 0.43, 0.50, 0.59, 0.67 \mbox{ TeV}, \\
  M_h &=& 0.42, 0.44, 0.46, 0.47 \mbox{ TeV}, \\
  M_H &=& 0.43, 0.50, 0.59, 0.67 \mbox{ TeV}, 
\end{eqnarray}
respectively.
Note that $\tan\beta \equiv v_{t'}/v_{b'} = 1$ in our model.
The $HZZ$-, $h\bar{t}'t'$- and $H\bar{t}'t'$-couplings are proportional 
to~\cite{Higgs-hunter}
\begin{align}
&  \cos(\beta-\alpha)=-0.02, -0.002, -0.0009, -0.0005, \\
&  \cos\alpha/\sin\beta=0.98, 1.0, 1.0, 1.0, \\
&  \sin\alpha/\sin\beta=-1.0, -1.0, -1.0, -1.0,
\end{align}
respectively, 
where $\alpha$ denotes the mixing angle of the two CP even 
Higgs bosons.
We can immediately read the relative 
$h\bar{b}'b'$- and $H\bar{b}'b'$-couplings, 
$-\sin\alpha/\cos\beta$ and $\cos\alpha/\cos\beta$, from above,
because of $\tan\beta=1$ in our model.
Due to $M_{H^\pm} \simeq M_H$ in this parameter regime,
the contributions of the Higgs bosons to 
the $S$- and $T$-parameters are small, 
at most $S_H = 0.03$ and $T_H=-0.05$ for the reference value of 
the SM Higgs boson $M_h^{ref}=300$ GeV.

Let us fix $m_0^{(3)}=1.0$ GeV and thereby obtain
\begin{eqnarray}
  \Delta g_t &=& \frac{m_0^{(3)}}{m_t}=
  (5.8 \pm 0.1) \times 10^{-3},\\
 \Delta g_b &=& \frac{m_0^{(3)}}{m_b} = 0.24
 \stackrel{+0.00}{\mbox{\scriptsize $-0.01$}},  
\end{eqnarray}
with the error bars.
The Higgs masses are estimated as
\begin{eqnarray}
  M_{\phi_t} &\simeq&
  \Lambda^{(3)} 
  \left(\frac{2\Delta g_t}{\ln \frac{1}{2\Delta g_t}}\right)^{1/2}
  = 0.051 \Lambda^{(3)}, \\
  M_{\phi_b} &\simeq&
  \Lambda^{(3)} 
  \left(\frac{2\Delta g_b}{\ln \frac{1}{2\Delta g_b}}\right)^{1/2}
  = 0.80 \Lambda^{(3)} ,  
\end{eqnarray}
where we used only the central value.
Recall that it is assumed in the present model
that the top-Higgs $\phi_t$ is decoupled.
It requires, say, $M_{\phi_t} \gtrsim 1$ TeV, i.e., 
$\Lambda^{(3)} \gtrsim 20$ TeV.
We also find
\begin{eqnarray}
 \Lambda^{(34)} &\simeq& \Lambda^{(4)}
 \sqrt{\frac{C_2 g_{t't}^2}{4\pi^2}\frac{m_{t'}}{m_0^{(3)}}}
 = 2.7 \sqrt{C_2} g_{t't} \Lambda^{(4)}, \\
 &\simeq& \Lambda^{(4)}
 \sqrt{\frac{C_2 g_{b'b}^2}{4\pi^2}\frac{m_{b'}}{m_0^{(3)}}}
 = 2.7 \sqrt{C_2} g_{b'b} \Lambda^{(4)} \, .   
\end{eqnarray}

For the masses of the 2nd family, 
assuming $\eta_t^{(23)} = \eta_b^{(23)} \equiv \eta^{(23)}$,
the following relation is crucial;
\begin{equation}
  \eta^{(23)} = \frac{m_c-m_s}{m_t-m_b}
 =(7.0 \stackrel{+0.7}{\mbox{\scriptsize $-0.9$}})\times 10^{-3},
\end{equation}
so that we obtain
\begin{eqnarray}
&&  m_0^{(2)} = m_s - \eta^{(23)} m_b
  = 75\stackrel{+30}{\mbox{\scriptsize $-39$}}\mbox{ MeV}, \\[2mm]
&&  C_2 g_{tc(bs)}^2 
    \frac{(\Lambda^{(34)})^2}{(\Lambda^{(23)})^2}
    \simeq 0.28 \stackrel{+0.03}{\mbox{\scriptsize $-0.04$}}, 
\end{eqnarray}
i.e.,
\begin{eqnarray}
  \Lambda^{(23)} &=& (1.9 \pm 0.1)\sqrt{C_2} g_{tc} \Lambda^{(34),}\\
                &=& (1.9 \pm 0.1)\sqrt{C_2} g_{bs} \Lambda^{(34)} \, .
\end{eqnarray}
The mass $m_0^{(2)}$ yields
\begin{align}
 \Lambda^{(24)} &\simeq \Lambda^{(4)}
 \sqrt{\frac{C_2 g_{t'c}^2}{4\pi^2}\frac{m_{t'}}{m_0^{(2)}}}
 = (10 \stackrel{+4}{\mbox{\scriptsize $-2$}})
   \sqrt{C_2} g_{t'c} \Lambda^{(4)}, \\
 &\simeq \Lambda^{(4)}
 \sqrt{\frac{C_2 g_{b's}^2}{4\pi^2}\frac{m_{b'}}{m_0^{(2)}}}
 = (10 \stackrel{+4}{\mbox{\scriptsize $-2$}})
   \sqrt{C_2} g_{b's} \Lambda^{(4)} \, .   
\end{align}
Finally, for the 1st family, we directly get
\begin{eqnarray}
 \Lambda^{(14)} &\simeq& \Lambda^{(4)}
 \sqrt{\frac{C_2 g_{t'u}^2}{4\pi^2}\frac{m_{t'}}{m_0^{(1)}}}
 = 61 \sqrt{C_2} g_{t'u} \Lambda^{(4)}, \\
 &\simeq& \Lambda^{(4)}
 \sqrt{\frac{C_2 g_{b'd}^2}{4\pi^2}\frac{m_{b'}}{m_0^{(1)}}}
 = 61 \sqrt{C_2} g_{b'd} \Lambda^{(4)} , 
\end{eqnarray}
where we used
\begin{equation}
  m_0^{(1)}=2.0\mbox{ MeV} \, .
\end{equation}
Note that in order to get a more realistic ratio,
\begin{equation}
  \frac{m_u}{m_d} = 0.35 - 0.60,
\end{equation}
one may tune $g_{t'u}^2/g_{b'd}^2$ to the latter. 
One should however remember that
the eigenvalues $m_u$ and $m_d$ are determined
after diagonalizing the quark mass matrices discussed in Sec.~\ref{2.3}.
In general, the numerical calculations of $m_u$ and $m_d$
beyond the order-estimates are highly sensitive to
the fine-structure of the mass matrices.

In summary, the suppression factors 
for getting the masses
$m_0^{(3)}=1.0$ GeV, $m_0^{(2)}=75$ MeV and $m_0^{(1)}=2.0$ MeV
should be equal to
\begin{eqnarray}
&&  \eta_{t'}^{(3)} = \frac{m_0^{(3)}}{m_{t'}}=3.4 \times 10^{-3},
    \quad
    \eta_{b'}^{(3)} = \frac{m_0^{(3)}}{m_{b'}}=3.4 \times 10^{-3}, 
    \nonumber \\ && \\
&&  \eta_{t'}^{(2)} = \frac{m_0^{(2)}}{m_{t'}}=2.6 \times 10^{-4},
    \quad
    \eta_{b'}^{(2)} = \frac{m_0^{(2)}}{m_{b'}}=2.6 \times 10^{-4}, 
    \nonumber \\ && \\
&&  \eta_{t'}^{(1)} = \frac{m_0^{(1)}}{m_{t'}}=6.8 \times 10^{-6},
    \quad
    \eta_{b'}^{(1)} = \frac{m_0^{(1)}}{m_{b'}}=6.9 \times 10^{-6}.
    \nonumber \\ && 
\end{eqnarray}
They are obtained by taking appropriate values for the ratios 
$\Lambda^{(i4)}/\Lambda^{(4)}$ as above.

\begin{table*}
  \centering
  \begin{tabular}{c|cc|cc|}\hline
   $\Lambda^{(4)}$ (TeV) & 30 & 30 & 20 & 20 \\ \hline
   $m_{t'}$ (TeV) & $0.26$ & $0.26$ & $0.27$ & $0.27$ \\
   $m_{b'}$ (TeV) & $0.26$ & $0.26$ & $0.27$ & $0.27$ \\  \hline
   $m_0^{(3)}$ (GeV) & $1.0$ & $2.0$ & $1.0$ & $2.0$ \\
   $\Delta g_t$ & $5.8 \times 10^{-3}$ & $0.012$ 
                & $5.8 \times 10^{-3}$ & $0.012$ \\
   $\Delta g_b$ & $0.24$ & $0.48$ & $0.24$ & $0.48$ \\
   $M_{\phi_t}/\Lambda^{(3)}$ & $0.051$ & $0.079$ & $0.051$ & $0.079$ \\
   $M_{\phi_b}/\Lambda^{(3)}$ & $0.80$ & $4.4$ & $0.80$ & $4.4$ \\
   $(\Lambda^{(34)})^2/[C_2 g_{q'q}^2 (\Lambda^{(4)})^2]$ 
   & $6.8$ & $3.2$ & $6.8$ & $3.6$ \\ \hline
   $m_0^{(2)}$ (MeV) & 
   $75\stackrel{+30}{\mbox{\scriptsize $-39$}}$ &
   $75\stackrel{+30}{\mbox{\scriptsize $-39$}}$ &
   $75\stackrel{+30}{\mbox{\scriptsize $-39$}}$ &
   $75\stackrel{+30}{\mbox{\scriptsize $-39$}}$ \\
   $\eta^{(23)}$ &
   $(7.0 \stackrel{+0.7}{\mbox{\scriptsize $-0.9$}})\times 10^{-3}$ &
   $(7.0 \stackrel{+0.7}{\mbox{\scriptsize $-0.9$}})\times 10^{-3}$ &
   $(7.0 \stackrel{+0.7}{\mbox{\scriptsize $-0.9$}})\times 10^{-3}$ &
   $(7.0 \stackrel{+0.7}{\mbox{\scriptsize $-0.9$}})\times 10^{-3}$ \\
   $(\Lambda^{(24)})^2/[C_2 g_{q'q}^2 (\Lambda^{(4)})^2]$ &
   $88 \stackrel{+96}{\mbox{\scriptsize $-25$}}$ &
   $88 \stackrel{+96}{\mbox{\scriptsize $-25$}}$ &
   $92 \stackrel{+99}{\mbox{\scriptsize $-27$}}$ &
   $92 \stackrel{+99}{\mbox{\scriptsize $-27$}}$ \\
   $(\Lambda^{(23)})^2/[C_2 g_{q q}^2 (\Lambda^{(34)})^2]$ &
   $3.6 \stackrel{+0.5}{\mbox{\scriptsize $-0.3$}}$ & 
   $3.6 \stackrel{+0.5}{\mbox{\scriptsize $-0.3$}}$ & 
   $3.6 \stackrel{+0.5}{\mbox{\scriptsize $-0.3$}}$ & 
   $3.6 \stackrel{+0.5}{\mbox{\scriptsize $-0.3$}}$ \\ \hline
   $m_0^{(1)}$ (MeV) & 1 & 2 & 1 & 2 \\
   $(\Lambda^{(14)})^2/[C_2 g_{q'q}^2 (\Lambda^{(4)})^2]$ & 
   $6.6 \times 10^3$ & $3.2\times 10^3$
   & $6.9 \times 10^3$ & $3.5 \times 10^3$ \\ \hline
  \end{tabular}
  \caption{Numerical estimates for $\Lambda^{(4)}=20,30$ TeV.}
\end{table*}

\begin{table*}
  \centering
  \begin{tabular}{c|cc|cc|}\hline
   $\Lambda^{(4)}$ (TeV) & 10 & 10 & 5 & 5 \\ \hline
   $m_{t'}$ (TeV) & $0.29$ & $0.29$ & $0.32$ & $0.32$ \\
   $m_{b'}$ (TeV) & $0.29$ & $0.29$ & $0.32$ & $0.32$ \\ \hline
   $m_0^{(3)}$ (GeV) & $1.0$ & $2.0$ & $1.0$ & $2.0$ \\
   $\Delta g_t$ & $5.8 \times 10^{-3}$ & $0.012$ 
                & $5.8 \times 10^{-3}$ & $0.012$ \\
   $\Delta g_b$ & $0.24$ & $0.48$ & $0.24$ & $0.48$ \\
   $M_{\phi_t}/\Lambda^{(3)}$ & $0.051$ & $0.079$ & $0.051$ & $0.079$ \\
   $M_{\phi_b}/\Lambda^{(3)}$ & $0.80$ & $4.4$ & $0.80$ & $4.4$ \\
   $(\Lambda^{(34)})^2/[C_2 g_{q'q}^2 (\Lambda^{(4)})^2]$ 
   & $7.3$ & $3.6$ & $8.4$ & $4.0$ \\ \hline
   $m_0^{(2)}$ (MeV) & 
   $75\stackrel{+30}{\mbox{\scriptsize $-39$}}$ &
   $75\stackrel{+30}{\mbox{\scriptsize $-39$}}$ &
   $75\stackrel{+30}{\mbox{\scriptsize $-39$}}$ &
   $75\stackrel{+30}{\mbox{\scriptsize $-39$}}$ \\
   $\eta^{(23)}$ &
   $(7.0 \stackrel{+0.7}{\mbox{\scriptsize $-0.9$}})\times 10^{-3}$ &
   $(7.0 \stackrel{+0.7}{\mbox{\scriptsize $-0.9$}})\times 10^{-3}$ &
   $(7.0 \stackrel{+0.7}{\mbox{\scriptsize $-0.9$}})\times 10^{-3}$ &
   $(7.0 \stackrel{+0.7}{\mbox{\scriptsize $-0.9$}})\times 10^{-3}$ \\
   $(\Lambda^{(24)})^2/[C_2 g_{q'q}^2 (\Lambda^{(4)})^2]$ &
   $99 \stackrel{+106}{\mbox{\scriptsize $-29$}}$ &
   $99 \stackrel{+106}{\mbox{\scriptsize $-29$}}$ &
   $109 \stackrel{+119}{\mbox{\scriptsize $-31$}}$ &
   $109 \stackrel{+119}{\mbox{\scriptsize $-31$}}$ \\
   $(\Lambda^{(23)})^2/[C_2 g_{q q}^2 (\Lambda^{(34)})^2]$ &
   $3.6 \stackrel{+0.5}{\mbox{\scriptsize $-0.3$}}$ & 
   $3.6 \stackrel{+0.5}{\mbox{\scriptsize $-0.3$}}$ & 
   $3.6 \stackrel{+0.5}{\mbox{\scriptsize $-0.3$}}$ & 
   $3.6 \stackrel{+0.5}{\mbox{\scriptsize $-0.3$}}$ \\ \hline
   $m_0^{(1)}$ (MeV) & 1 & 2 & 1 & 2 \\
   $(\Lambda^{(14)})^2/[C_2 g_{q'q}^2 (\Lambda^{(4)})^2]$ & 
   $7.4 \times 10^3$ & $3.7 \times 10^3$ & 
   $8.3 \times 10^3$ & $4.1 \times 10^3$ \\ \hline
  \end{tabular}
  \caption{Numerical estimates for $\Lambda^{(4)}=5,10$ TeV.}
\end{table*}

Roughly speaking, in our scenario, the masses of $t'$ and $b'$ are
\begin{equation}
  m_{t'} \simeq 300\mbox{ GeV}, \quad m_{b'} \simeq 300\mbox{ GeV}, 
\end{equation}
the cutoff scale is
\begin{equation}
  \Lambda^{(4)} \sim 10\mbox{ TeV},
\end{equation}
and the other FCN scales are estimated as 
\begin{align}
&  \Lambda^{(14)} \sim 100 \Lambda^{(4)}, 
&  \Lambda^{(24)} \sim 10 \Lambda^{(4)} , \\
&  \Lambda^{(34)} \sim 3 \Lambda^{(4)}, 
&  \Lambda^{(23)} \sim 5 \Lambda^{(4)}\,.
\end{align}
Although the exchange of $\Lambda^{(34)}$ contributes to $R_b$,
it is tiny, $\delta R_b/R_b \sim 10^{-6}$ for 
$\Lambda^{(34)}=30$ TeV with $C_2 g_{b'b}^2 \sim 1$.
The constraint from the $B_s^0$-$\bar{B}_s^0$ mixing suggests
$\Lambda^{(23)} \gtrsim 100$ TeV, so that 
the above estimate $\Lambda^{(23)} \sim 5 \Lambda^{(4)}$ 
is a bit dangerous.
(If we take a smaller $m_0^{(3)}$ or a bigger $\Lambda^{(4)}$, 
we can evade this problem.)

As we discussed in Sec.~\ref{2.3}, 
the constraints from the $B^0$--$\bar{B}^0$ mixing, $ b \to s \gamma$
and $R_b$ via the $t'$ loop are suppressed, 
because the relevant mixing angles are tiny, 
$|V_{t'd}| \sim |V_{cd}| m_c/m_{t'} \sim 10^{-3}$, and 
$|V_{t's}| \simeq |V_{t'b}| \sim m_c/m_{t'} \sim 10^{-2}$.
The contributions of the charged Higgs are also suppressed.

The numerical estimates of all the relevant parameters of 
the model for the values $\Lambda^{(4)}= 30$ TeV, 
$\Lambda^{(4)}= 20$ TeV, and $\Lambda^{(4)}= 10$ TeV, 
$\Lambda^{(4)}= 5$ TeV are presented in Tables I and II, respectively.

The following comments are in order.
(i) While the contribution of the particles
of the 4th family to
the $T$-parameter is almost vanishing in the 
case of degenerate masses of both 
the quarks and the leptons,
their contribution to the $S$-parameter is a bit large,
$S_f \sim 0.2$, if no Majorana neutrinos are present.
One can avoid this difficulty by introducing a Majorana
neutrino with a mass smaller than that of 
the charged lepton~\cite{Gates:1991uu,Appelquist:1993gi}.
At the same time, the 
$T$-parameter can be kept small even in 
this case~\cite{Bertolini:1990ek,Gates:1991uu,Holdom:1996bn}.
(ii) In the present model, the maximum value for the mass of $t'$ and $b'$
is realized for $\Lambda^{(4)} = m_{t'(b')}$.
The PS formula (\ref{PS}) yields $m_{t'(b')}^{({\rm max})} \simeq 1$ TeV 
for it.
The fact that $m_{t'} \simeq m_{b'} < 1$ TeV in this model is
noticeable: the 4th family quarks with masses of 1 TeV or lighter
can be discovered at LHC~\cite{AguilarSaavedra:2005pv}.

\section{Discussion}
\label{4}

The two crucial ingredients in the class of models described in 
this paper
are (i) the assumption that the EWSB dynamics leads to the isospin symmetric
quark mass spectrum, with the masses of the order of the down-type quarks, and
(ii) the existence of strong (although subcritical) horizontal 
diagonal interactions
for the $t$ quark plus horizontal flavor-changing neutral interactions 
between different families. The signature of such dynamics is the presence of
composite Higgs bosons. It is noticeable that this dynamics can be
build into the scenarios with different EWSB mechanisms.

The concrete model with the 4th family considered above
shows that these two ingredients quite naturally lead to the realistic masses
for quarks. 
Moreover, as was pointed out in Sec.~\ref{2}, 
in the present approach it is necessary
to choose the mass $m_0^{(2)}$ (generated by the EWSB dynamics)
to be of the order of the mass of the $s$ quark: only in this 
case one can obtain the correct $m_c/m_s$ ratio.
We also demonstrated that
by using a simple extension of
the present mechanism for producing the quark masses, 
the essential features of the CKM matrix can be extracted.
Another noticeable feature in the model is the absence of fine tuning:
the nearcriticality (1 part in $10^2$) of the coupling of the $t$ quark is 
determined by the small ratio $m_b/m_t \simeq 2.5 \times 10^{-2}$.
 
As the next steps, it would be important to include leptons
and to study 
the dynamics underlying the CKM matrix in more detail. As to the leptons,
the fact that the masses of the charged leptons are of the order of the masses
of the corresponding down-type quarks
suggests that it is not unreasonable to assume
that the origin of the former is similar to that of the latter.
The main specific issues for leptons are of course connected with neutrinos:
in particular, with a large mixing between the muon and tau neutrinos and
a possible existence of Majorana neutrinos. Note that the latter occur 
quite naturally in the 4th family models~\cite{4family}.
Last but not least, it would be interesting to embed the present scenario into
an extra dimensional one~\cite{Dobrescu:1998dg,Burdman:2007sx}.

\acknowledgments

This work was supported by the Natural Sciences and Engineering Research
Council of Canada.

\appendix

\section{More about isospin symmetry breaking in the third family}
\label{A}

In this section, we will briefly describe 
several models which could provide strong isospin symmetry breaking 
in the third family. 

For example, we here employ the first version of 
the topcolor model~\cite{Hill:1991at,Hill:2002ap}. 
In this case, the QCD sector in the SM is extended to a  
$SU(3)_1 \times SU(3)_2$ one, with a stronger coupling for the 
$SU(3)_1$. The $SU(3)_1$ and $SU(3)_2$ charges are assigned as 
\begin{align}
&  (u,d)_L \to ({\bf 1}, {\bf 3}), &
    & u_R \to ({\bf 1}, {\bf 3}) ,  & d_R \to ({\bf 1}, {\bf 3}) ,  \\
&  (c,s)_L \to ({\bf 1}, {\bf 3}),  &
    & c_R \to ({\bf 1}, {\bf 3}) ,  & s_R \to ({\bf 1}, {\bf 3}) ,  \\
&  (t,b)_L \to ({\bf 3}, {\bf 1}),  &
    & t_R \to ({\bf 3}, {\bf 1}),   & b_R \to ({\bf 1}, {\bf 3}) , \\
&  (t',b')_L \to ({\bf 3}, {\bf 1}), &
    & t'_R \to ({\bf 3}, {\bf 1}),   & b'_R \to ({\bf 3}, {\bf 1}) , 
\end{align}
while their $SU(2)_L \times U(1)_Y$ charges are conventional. 
Recall also that for the anomaly cancellation, 
$SU(2)_L$ singlet fermions are required,
\begin{equation}
  Q_L \to ({\bf 1}, {\bf 3}), \quad Q_R \to ({\bf 3}, {\bf 1}),
\end{equation}
with the same hypercharge as $b_R$~\cite{Hill:1991at}. 
Besides this topcolor scheme, 
we also introduce an additional $U(1)_{4F}$ gauge boson which couples 
(with the same strength) only to the fourth family.
We may assign the $U(1)_{4F}$ charge as the $U(1)_{B-L}$ one, for example.

Then, after the spontaneous breakdown of 
$SU(3)_1 \times SU(3)_2$ down to $SU(3)_c$
at the scale $\Lambda^{(3)}$ ($=\Lambda^{(4)}$ in this case), 
the NJL couplings for the top and bottom are 
$g_c^2 \cot^2\theta/(\Lambda^{(3)})^2$ and
$g_c^2/(\Lambda^{(3)})^2$, respectively,
where $g_c$ represents the QCD coupling constant
and $\theta$ is the mixing angle of the $SU(3)_{1,2}$ gauge bosons.
Since, unlike the topcolor model, we utilize the subcritical dynamics,
the following relation holds,
\begin{equation}
  \frac{3}{2\pi}\cot^2\theta \; \alpha_c(\Lambda^{(3)}) \lesssim 1.
\end{equation}
Therefore Eq.~(\ref{diff-t-b}) now reads 
\begin{equation}
  \frac{3}{2\pi}(\cot^2\theta - 1)\alpha_c(\Lambda^{(3)}) 
  \simeq \frac{m_0^{(3)}}{m_b},
\end{equation}
where $\alpha_c(\Lambda^{(3)})=g_c^2/(4\pi)$ is the QCD coupling
at the scale $\Lambda^{(3)}$.

As to $t'$ and $b'$, 
in order to make their NJL couplings supercritical, 
the contribution from $U(1)_{4F}$ is crucial,
\begin{equation}
    \frac{3}{2\pi}\cot^2\theta \; \alpha_c(\Lambda^{(4)}) 
  + \frac{3}{2\pi} \alpha_{4F}(\Lambda^{(4)})  \gtrsim 1.
\end{equation}
where $\alpha_{4F}(\Lambda^{(4)})=g_{4F}^2/(4\pi)$ is 
the gauge coupling of $U(1)_{4F}$ at the scale 
$\Lambda^{(4)}(=\Lambda^{(3)})$.

In this case, the scenario with three Higgs doublets 
as the composite fields of $t'$, $b'$ and $t$ is likely.

For a model with $\Lambda^{(3)} \gtrsim \Lambda^{(4)}$,
we may further extend the QCD sector,
\begin{equation}
  SU(3)_1 \times SU(3)_{2+b} \times SU(3)_t \times SU(3)_4,
\end{equation}
with the following quark representations:
\begin{widetext}
\begin{eqnarray}
 & (u,d)_L \to ({\bf 3}, {\bf 1}, {\bf 1}, {\bf 1}), \quad 
    u_R \to ({\bf 3}, {\bf 1}, {\bf 1}, {\bf 1}), \quad 
    d_R \to ({\bf 3}, {\bf 1}, {\bf 1}, {\bf 1}), \\
 & (c,s)_L \to ({\bf 1}, {\bf 3}, {\bf 1}, {\bf 1}), \quad 
    c_R \to ({\bf 1}, {\bf 3}, {\bf 1}, {\bf 1}), \quad 
    s_R \to ({\bf 1}, {\bf 3}, {\bf 1}, {\bf 1}), \\
 & (t,b)_L \to ({\bf 1}, {\bf 1}, {\bf 3}, {\bf 1}), \quad 
    t_R \to ({\bf 1}, {\bf 1}, {\bf 3}, {\bf 1}), \quad 
    b_R \to ({\bf 1}, {\bf 3}, {\bf 1}, {\bf 1}), \\
 & (t',b')_L \to ({\bf 1}, {\bf 1}, {\bf 1}, {\bf 3}), \quad 
    t'_R \to ({\bf 1}, {\bf 1}, {\bf 1}, {\bf 3}), \quad 
    b'_R \to ({\bf 1}, {\bf 1}, {\bf 1}, {\bf 3}) \, .
\end{eqnarray}
\end{widetext}
The charges of the SM gauge group $SU(2)_L \times U(1)_Y$
are conventional.
For anomaly cancellation, we also introduce
$SU(2)_L$--singlet quarks,
\begin{equation}
  Q_L \to ({\bf 1}, {\bf 3}, {\bf 1}, {\bf 1}), \quad
  Q_R \to ({\bf 1}, {\bf 1}, {\bf 3}, {\bf 1}),
\end{equation}
with the same hypercharge as $b_R$.

At the scale $\Lambda^{(3)}$, a part of the gauge symmetry is 
spontaneously broken down to a diagonal subgroup,
\begin{equation}
  SU(3)_{2+b} \times SU(3)_t \to SU(3)', 
\end{equation}
and also, at the scale $\Lambda^{(4)}$,
the rest part is broken down to
\begin{equation}
    SU(3)_1 \times SU(3)_4 \to SU(3)'' \, . 
\end{equation}
The two gauge groups are broken down to the conventional QCD
at some scale $\Lambda_c$ ($\sim \Lambda^{(4)}$),
\begin{equation}
  SU(3)' \times SU(3)'' \to SU(3)_c \, .
\end{equation}
The gauge coupling constants then satisfy the following relations,
\begin{equation}
  \frac{1}{g_{2c}^2}+\frac{1}{g_{tc}^2}=\frac{1}{g'_c{}^2}, \quad
  \frac{1}{g_{1c}^2}+\frac{1}{g_{4c}^2}=\frac{1}{g''_c{}^2}, 
\end{equation}
and 
\begin{equation}
    \frac{1}{g'_c{}^2} + \frac{1}{g''_c{}^2} = \frac{1}{g_c^2},
\end{equation}
where $g_{ic}$ ($i=1,2,t,4$), $g'_c$ and $g''_c$
denote the gauge couplings for $SU(3)_{1,(2+b),t,4}$, $SU(3)'$ and
$SU(3)''$, respectively.
Let us introduce the mixing angles $\theta'_c$, $\theta''_c$ and 
$\theta_c$ between $SU(3)_{2+b}$ and $SU(3)_t$, between
$SU(3)_1$ and $SU(3)_4$, and between $SU(3)'$ and $SU(3)''$, 
respectively.
At the scale $\Lambda^{(3)}$, the four-top interaction is generated
with the strength 
\begin{equation}
 G_t \equiv g'_c{}^2 \cot^2\theta'_c/(\Lambda^{(3)})^2, 
\end{equation}
whereas the strengths of the NJL interactions for $t'$ and $b'$ are 
\begin{equation}
 G_4 \equiv g''_c{}^2 \cot^2\theta''_c/(\Lambda^{(4)})^2,  
\end{equation}
provided at the scale $\Lambda^{(4)}$.
When $g'_c \sim g''_c \sim g_c$, i.e., $\tan\theta_c \sim 1$,
the four-fermion interactions generated at the scale $\Lambda_c$
are irrelevant.
In our scenario, we require that $G_t$ and $G_4$ are subcritical
and supercritical, respectively,
so that 
\begin{equation}
 \frac{3}{2\pi} \frac{\cot^2\theta'_c}{\sin^2\theta_c}
 \alpha_c(\Lambda^{(3)}) \lesssim 1, 
\end{equation}
at $\Lambda^{(3)}$, and
\begin{equation}
  \frac{3}{2\pi} \frac{\cot^2\theta''_c}{\cos^2\theta_c}
  \alpha_c(\Lambda^{(4)})\gtrsim 1,
\end{equation}
at $\Lambda^{(4)}$, where we expressed the gauge couplings 
$g'_c$ and $g''_c$ through $g_c$ and the mixing angle $\theta_c$,
i.e., $g'_c = g_c/\sin\theta_c$ and $g''_c = g_c/\cos\theta_c$.
Note that the NJL couplings for the 3rd family are restricted by
the current mass enhancement relations.
Eq.~(\ref{diff-t-b}) then reads 
\begin{equation}
  \frac{3}{2\pi}\frac{\cot^2\theta'_c - 1}{\sin^2\theta_c}
  \alpha_c(\Lambda^{(3)}) 
  \simeq \frac{m_0^{(3)}}{m_b} \, .
\end{equation}

Another possibility for the isospin symmetry breaking is to use
the $U(1)$-tilting mechanism, which can be realized in the model
with extended QCD and hypercharge sectors,
$SU(3)_1 \times SU(3)_2 \times U(1)_1 \times U(1)_2$~\cite{Hill:1994hp,Lane:1995gw}.

In this paper, we did not discuss the origin of the FCN interactions.
For such a purpose, concrete ETC models could provide 
a useful hint~\cite{Appelquist:1993gi,Appelquist:1993sg,Appelquist:2003hn}.

\end{document}